\documentclass[twocolumn,preprintnumbers,amsmath,amssymb]{revtex4}
 \usepackage{graphicx}

\begin{document}
 \preprint{}

 \title{Preparation of an ultra-cold sample of ammonia molecules for precision measurements}

 \author{Marina Quintero-P\'{e}rez$^{1}$}
 \author{Thomas E. Wall$^{1}$\footnote{Present address: Department of Physics and Astronomy, University College London, Gower Street, London WC1E 6BT, UK}}
 \author{Steven~Hoekstra$^{2}$}
 \author{Hendrick L.~Bethlem$^{1}$}
\email{H.L.Bethlem@vu.nl}

\affiliation{
$^{1}${LaserLaB, Department of Physics and Astronomy, VU University Amsterdam, De Boelelaan 1081, 1081 HV, Amsterdam, The Netherlands}\\
$^{2}${University of Groningen, Faculty of Mathematics and Natural Sciences, Zernikelaan 25, 9747 AA, Groningen, The Netherlands}
}

 \begin{abstract}
We present experiments in which an ultra-cold sample of ammonia molecules is released from an electrostatic trap and recaptured after a variable time. It is shown that, by performing adiabatic cooling before releasing the molecules and adiabatic re-compression after they are recaptured, we are able to observe molecules even after more than 10\,ms of free expansion. A coherent measurement performed during this time will have a statistical uncertainty that decreases approximately as the inverse of the square root of the expansion time. This offers interesting prospects for high-resolution spectroscopy and precision tests of fundamental physics theories.   
 
 \end{abstract}

\maketitle
 
\section{Introduction}

After eliminating more mundane broadening effects, the resolution of a spectroscopic measurement is ultimately limited by the time an atom or molecule interacts with the radiation field. In typical molecular beam experiments, this interaction time is limited to around 1 ms, resulting in a measured linewidth on the order of a kilohertz. The development of cooling techniques for atoms has led to a dramatic increase in the interaction times and hence the attained accuracy. Currently, atom and ion clocks reach accuracies on the order of 10$^{-18}$, allowing for tests of fundamental physics theories such as relativity~\cite{Chou:Science2010}, quantum electrodynamics~\cite{Hansch2006} and the time-invariance of the fine-structure constant~\cite{Rosenband2008} at an unprecedented level. It is anticipated that cooling techniques for molecules will lead to a similar increase in accuracy, allowing for more precise tests of time-reversal symmetry~\cite{Hudson:Nature2011,ACME2014} and the time-invariance of the proton-to-electron mass ratio~\cite{Shelkovnikov2008}, observation of parity violation~\cite{DeMille:PRL2008} and weak interactions in chiral molecules~\cite{Daussy1999}. Unfortunately, the cooling techniques for molecules demonstrated so far, such as Stark and Zeeman deceleration, buffer gas cooling, photo-association and laser cooling, have a poor efficiency (for recent review papers, see Refs.~\cite{Carr:NJP2009,vandeMeerakker:ChemRev2012,Lemeshko2013}). Therefore, although in proof-of-principle experiments prolonged interaction times have been achieved~\cite{Veldhoven2004, Hudson:PRL2006}, this was accomplished at the expense of a large decrease in signal. Hence, the Allan deviation -- a measure of the statistical uncertainty of the experiment -- was increased rather than decreased by using slow molecules. But the statistical uncertainty is not all that matters, usually the accuracy of experiments is limited by systematic effects rather than by the statistical uncertainty. As most systematic effects are proportional to the measured line-width~\cite{Amy-Klein1999}, an experiment with a large Allan deviation but a small line-width is in some cases preferable over an experiment with a smaller Allan deviation but a large line-width. If the Allan deviation is too large, however, the required measurement times become unrealistically long.  

Recently, our group has decelerated and trapped ammonia molecules using a series of rings to which oscillating voltages are applied~\cite{Quintero-Perez2013, Jansen2013}. The advantage of such a traveling-wave decelerator~\cite{Meek2008,Osterwalder2010,Meek2011} is that molecules are confined in a genuine 3D trap throughout the deceleration process, which avoids the losses at low velocity that plague conventional Stark decelerators. The necessary voltages for the decelerator are generated by amplifying the output of an arbitrary wave generator using fast high voltage (HV) amplifiers, giving us great control over the trapping potential. In this paper, we present measurements in which we adiabatically cool ND$_{3}$ molecules to sub-mK temperatures, release them by switching off the trap voltages completely, recapture them after a variable expansion time, re-compress and finally detect them. We discuss how this method could be used for obtaining high-resolution spectra.

\section{Experimental Setup}

\begin{figure}[h!]
\includegraphics{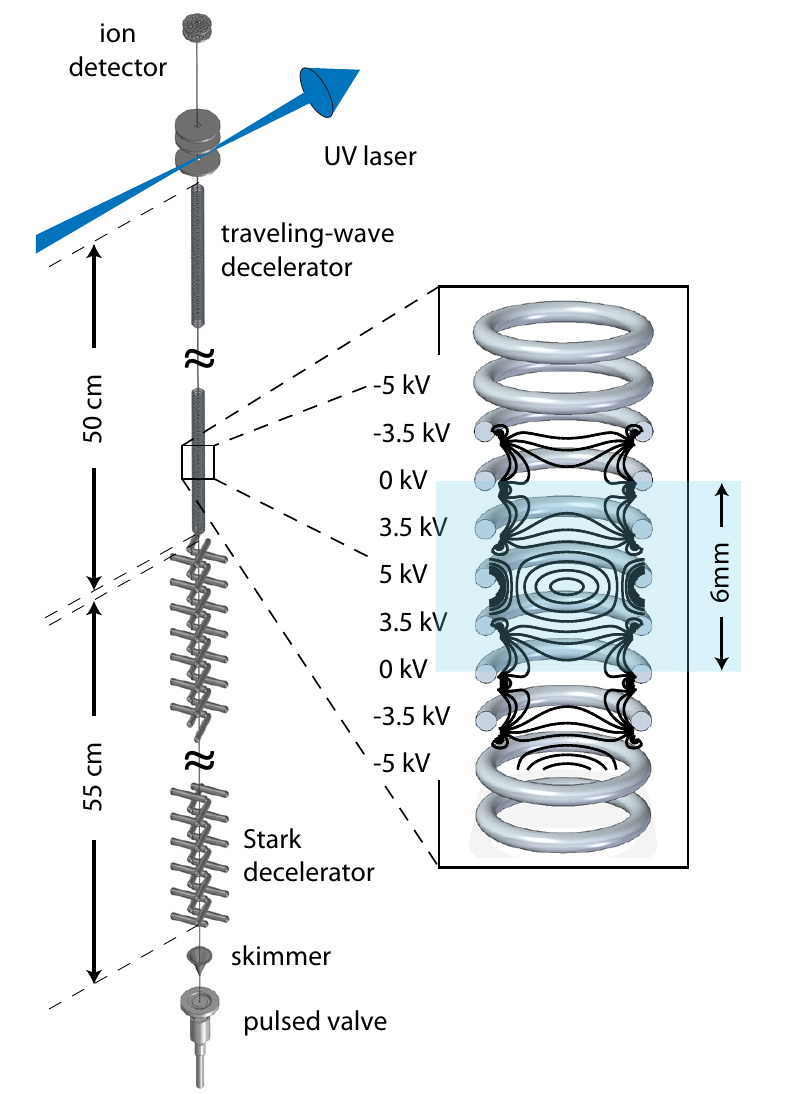}
\caption{\label{fig1-Setup}
Schematic view of the vertical molecular beam machine. A beam of molecules is decelerated and trapped using a combination of a Stark decelerator and a traveling-wave decelerator. The inset shows the electric field magnitude inside the  traveling-wave decelerator (in steps of 2.5\,kV/cm) with the voltages as indicated.}
\end{figure}

Fig.~\ref{fig1-Setup} shows a schematic view of our vertical molecular beam machine. A beam of ammonia molecules seeded in xenon is released into vacuum using a pulsed valve. This beam is decelerated from around 330\,m/s to 90\,m/s using a Stark decelerator consisting of 101 pairs of electrodes to which voltages of +10 and ${-}$10 kV are applied. The voltages are abruptly ($<$~100\,ns) switched using four independent HV switches (Behlke HTS 151-03-GSM) that are triggered by a programmable delay generator. The traveling-wave decelerator, which is mounted 24 mm above the last electrode pair of the first decelerator, consists of a series of 336 rings, each of which is attached to one of eight stainless steel rods to which voltages of up to $\pm$5 kV are applied. At any moment in time, the voltages applied to successive ring electrodes follow a sinusoidal pattern in $z$, where $z$ is the position along the beam axis. These voltages create minima of electric field every 6\,mm, which act as 3D traps for weak-field seeking molecules. By modulating the voltages sinusoidally in time, the traps are moved along the decelerator. A modulation frequency that is constant in time results in a trap that moves with a constant positive velocity along the molecular beam axis. A constant acceleration or deceleration can be achieved by applying a linear chirp to the frequency~\cite{Osterwalder2010,Meek2011}. The necessary voltages are generated by amplifying the output of an arbitrary wave generator (Wuntronic DA8150) using eight fast HV amplifiers (Trek 5/80). The ND$_{3} $ molecules are state-selectively ionized 20\,mm above the traveling-wave decelerator using a focused UV laser beam and counted by an ion detector. More details of the setup are given in Quintero-P\'{e}rez \emph{et al.}~\cite{Quintero-Perez2013} and Jansen \emph{et al.}~\cite{Jansen2013}. Compared to the setup used in our previous experiments, we have made a number of changes: (i) the general valve has been replaced by a Jordan valve; (ii) the distance between the decelerator and the detection region has been decreased from 40\,mm to 20\,mm; (iii) the suspension of the traveling-wave decelerator has been altered such that the detection region is better pumped; (iv) the alignment of the ring electrodes with respect to each other as well as the alignment of the complete traveling-wave decelerator with respect to the beam line has been improved.

\section{Experimental Results}

\begin{figure}[h!]
  \centering
    \includegraphics{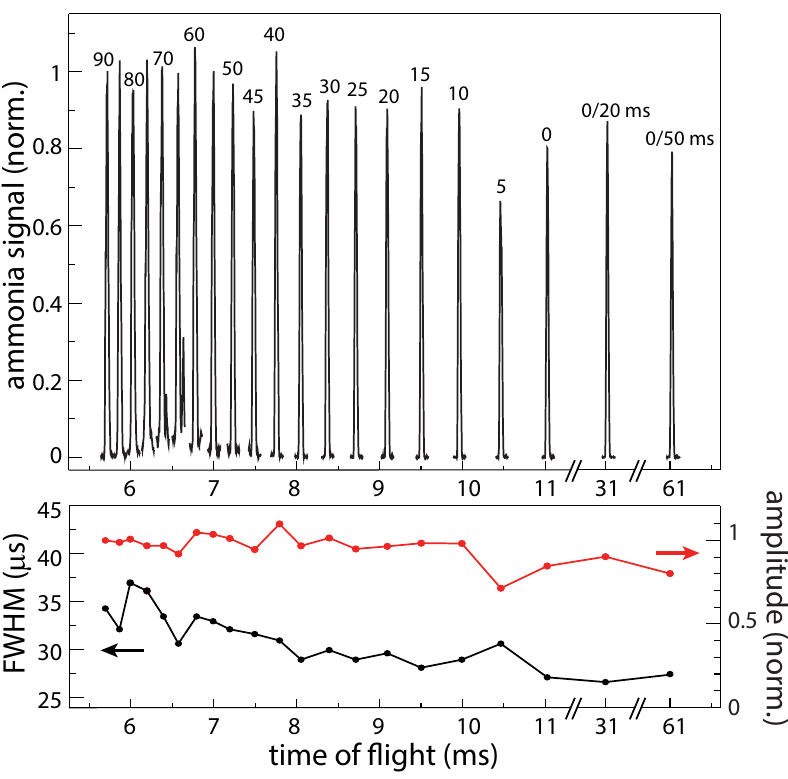}
  \caption{Measured time-of-flight (TOF) profiles when ND$_3$ molecules are guided at 90\,m/s, decelerated to 85-0\,m/s and trapped for 20 and 50\,ms before being accelerated back to 90\,m/s and detected (upper graph). The lower graph shows the full-width at half-maximum (FWHM)(black curve, left-hand side axis) and the amplitudes (red curve, right-hand side axis) of Gaussian fits to the TOF profiles.}
\label{fig2-TOFs}
\end{figure}

The upper panel of Fig.~\ref{fig2-TOFs}  shows the density of ND$_{3}$ molecules above the decelerator as a function of time when different waveforms are applied to the ring decelerator. In all cases, a packet of molecules is decelerated to 90\,m/s using the conventional Stark decelerator and injected into the traveling-wave decelerator at $t=0$. The time-of-flight (TOF) trace labeled as `90' is recorded when a sinusoidal voltage with an amplitude of 5\,kV and a constant frequency of 7.5\,kHz is applied to the traveling-wave decelerator. In this case, molecules at a velocity of 90\,m/s are guided through the decelerator and arrive at the detection zone after 5.5\,ms. The TOF traces that are labeled as `85' to `0' are recorded when the frequency is first linearly decreased to a certain value and subsequently increased to its original value. With these waveforms, molecules are decelerated to the indicated value (in m/s) before being accelerated back to their original velocity. Consequently, the molecules will spend a longer time in the decelerator and will arrive at the detection zone at later times. The TOF trace that is labeled as `0' corresponds to the situation in which molecules come to a complete standstill before being accelerated back to 90\,m/s and detected. As expected, the time the molecules spend in the decelerator is now approximately 2 times longer than when the molecules are guided. In these experiments the acceleration is changed from 0\,m/s$^2$, for guiding, to 17000\,m/s$^2$, for decelaration to 0\,m/s. Once the molecules come to a standstill, we can hold them for arbitrarily long times by keeping the voltages constant. The profiles that are labeled as `0/20' and `0/50' correspond to the situations in which molecules are trapped for 20 and 50\,ms, respectively.

The lower panel of Fig.~\ref{fig2-TOFs} shows the full-width at half-maximum (FWHM)(black curve, left-hand side axis) and the amplitudes (red curve, right-hand side axis) of Gaussian fits to the TOF profiles. Since we detect molecules very close to the decelerator, the width of the TOF profiles mainly reflects the position spread of the guided molecules, i.e. the size of the trap in the $z$ direction. The width of the TOF profile measured when the molecules are guided corresponds to a position spread of 3\,mm. Deceleration of the molecules comes at the expense of the acceptance of the trap, which explains why the width of the TOF profiles decreases when the molecules are decelerated to lower velocities. The amplitudes of the Gaussian fits to the TOFs are normalized to that of the guided beam. As observed, the amplitude of the TOF profile measured after 50\,ms of trapping is about 20\% smaller than that of the guided beam. Note that in previous experiments~\cite{Quintero-Perez2013,Jansen2013} the ND$_{3}$ signal was reduced by 80$\%$, which was attributed to mechanical misalignments in the traveling-wave decelerator and a mismatch between the conventional Stark decelerator and the traveling-wave decelerator. These issues have been resolved by realigning the decelerator. In addition, the various improvements have resulted in an increase of the signal of the guided molecules of a factor of 12. 

It is observed in Fig.~\ref{fig2-TOFs} that the amplitude of the TOF profile measured when molecules are decelerated to 5\,m/s is smaller than expected while its width is larger than expected. We have repeated this measurement several times to ensure that it is not a statistical fluctuation, but this seems not to be the case. We attribute the observed heating to parametric amplification of the transverse motion when the transverse trapping frequency becomes similar to the inverse of the time that it takes molecules to move from one ring to the next. For ND$_{3}$ strong, broad resonances are expected at 1.9 and 3\,kHz, corresponding to a velocity of 2.8 and 4.5\,m/s \cite{Quintero-Perez2013}. These resonances do not lead to significant losses when the molecules are decelerated to a standstill -- apparently because the resonance is passed sufficiently fast. When molecules are decelerated to 5\,m/s, however, they spend a longer time near the resonance leading to considerable heating.         

\begin{figure}[h!]
  \centering
    \includegraphics{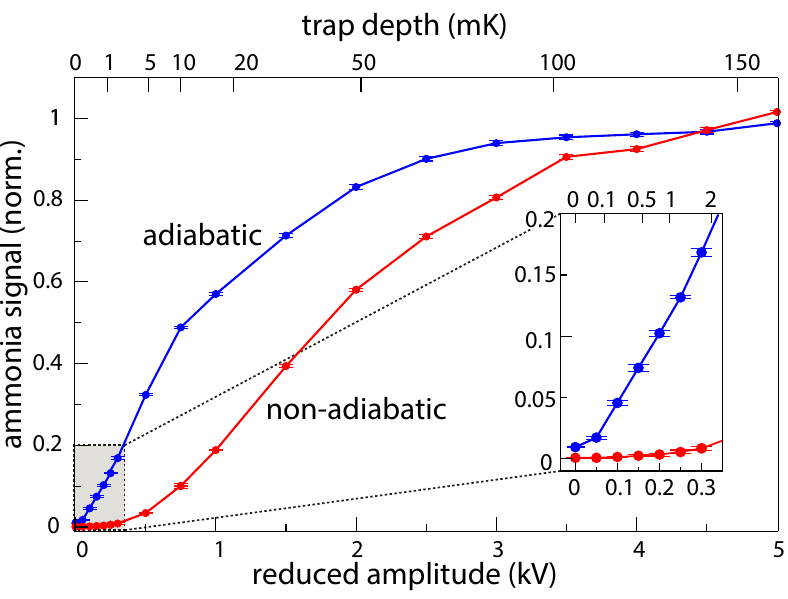}
  \caption{Trapped  ND$_{3}$ signal as a function of the amplitude of the waveform when, after deceleration and trapping, the voltages are slowly (blue curve labeled as `adiabatic') or abruptly (red curve labeled as `non-adiabatic') reduced to the indicated value. Each data point is the average of 600 shots; the error bars represent the one sigma statistical spread. The top axis shows the trap depth (in millikelvin) with this amplitude.}
\label{fig3-cooling}
\end{figure}

As the waveforms applied to the decelerator are generated by a computer code and loaded into an arbitrary wave generator, we have complete control over the shape and depth of the trap. This is illustrated by the measurements presented in Fig.~\ref{fig3-cooling}. In these measurements ND$_{3}$  molecules are again decelerated, trapped for a period of over 50\,ms, and subsequently accelerated and detected. In this case, however, while the molecules are trapped, the voltages applied to the decelerator are ramped down in 10\,ms, kept at a lower value for another 10\,ms and then ramped back up to 5\,kV over 10 ms (blue data points). Lowering the voltages of the trap has two effects: (i) the trap frequency is lowered, adiabatically cooling the molecules; (ii) the trap depth is reduced, allowing the hottest molecules to escape the trap. The bottom axis of Fig.~\ref{fig3-cooling} shows the value to which the amplitude of the waveform is reduced, while the top axis shows the resulting trap depth in mK. For comparison, the signal is also shown when the trap voltages are abruptly (10\,$\mu$s) lowered (red data points). In this case, no adiabatic cooling occurs and the signal decreases rapidly when the voltages are ramped down. The data presented in Fig.~\ref{fig3-cooling} is similar to that already presented in earlier publications~\cite{Quintero-Perez2013,Jansen2013}. However, as our signal is now much better than before, we can observe molecules even if the amplitude of the waveform is reduced to below 0.5\,kV,  as shown in the inset of the figure. It is seen that by reducing the trap depth to 200\,$\mu$K we still have more than 5\% of our signal left.

As discussed, the main motivation for our research on cold molecules is their potential for high-resolution spectroscopy. Ideally, one would like to perform spectroscopy on trapped molecules, for instance, by driving a transition between two states that have a similar Stark shift. However, a trap depth as low as 200\,$\mu$K already corresponds to a Stark shift of 4\,MHz. Thus, in order to measure this transition with high accuracy, the fractional difference between the Stark shifts of the levels involved should be very small. A more straight-forward approach is to switch off the trapping voltages completely. 

\begin{figure}[h!]
  \centering
    \includegraphics{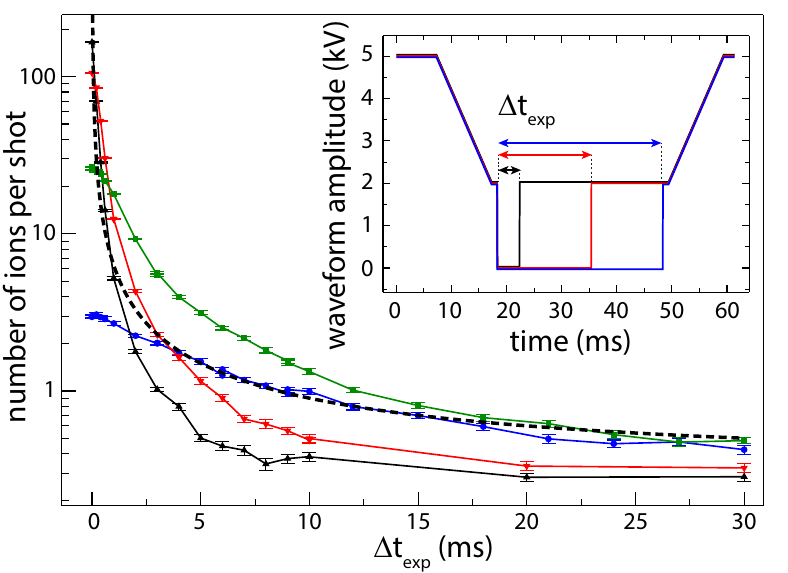}
  \caption{Number of ND$_{3}$ molecules detected after release and recapturing as a function of the time that the molecules are allowed to expand. Example waveform amplitudes are shown in the inset. The black curve is recorded when no adiabatic cooling or re-compression is performed, while for recording the red, green and blue curves the waveform amplitudes are first slowly reduced to 2, 0.5 and 0.1\,kV, respectively, before the molecules are released, and subsequently increased back from this value to 5\,kV after the molecules have been recaptured. The dashed black curve shows a 1/$\Delta t_{\mathrm{exp}}$ function to guide the eye.}
\label{fig4-recapturing}
\end{figure}

In Fig.~\ref{fig4-recapturing}, we show the results of an experiment in which the trap is switched off for a variable period of time. The waveforms used in this experiment, shown in the inset of the figure, are similar to the ones used for obtaining the data of Fig.~\ref{fig3-cooling}. In this case, however, after the voltages have been reduced to a certain value, $V_{\mathrm{red}}$, we abruptly switch off the voltages for a variable period of time, $\Delta t_{\mathrm{exp}}$. The $y$-axis shows the ion-current converted into the number of ions per shot on a logarithmic scale while the $x$-axis shows the expansion time in ms. The red, green and blue curves show the signal when the voltages are reduced to 2, 0.5 and 0.1\,kV, respectively. The black curve is recorded when no adiabatic cooling has been performed. As observed, without adiabatically cooling, the signal is initially large, but it drops rapidly as the expansion time is increased. When the molecules are adiabatically cooled before they are released and adiabatically recompressed after they are recaptured, the initial signal is smaller but it decays more slowly as $\Delta t_{\mathrm{exp}}$ is increased. Consequently, molecules are detected even for expansion times over 10\,ms. To understand the observations it is important to realize that our signal is proportional to the number of molecules in the laser focus and \emph{not} to the total number of molecules that are recaptured. For long expansion times, the number of molecules that are recaptured drops as 1/$\Delta t_{\mathrm{exp}}^3$; their density after re-compression, on the other hand, will stay approximately constant. As phase space density is conserved, this implies that the size of the trapped cloud shrinks. As our laser beam is tightly focused along the $y$ and $z$-direction, but elongated along the $x$-direction, expected to be proportional to 1/$\Delta t_{\mathrm{exp}}$, as is indeed observed. The dashed black curve in Fig.~\ref{fig3-cooling} shows a 1/$\Delta t_{\mathrm{exp}}$ function to guide the eye.

For the sake of simplicity, we have chosen the voltage used for adiabatic cooling to be the same as the one used to recapture the molecules. However, it is probably better to optimise these separately -- the voltage for adiabatic cooling should be chosen such that it satisfies the compromise between maximum cooling with a minimal loss of molecules, whereas the voltage applied to the trap when the molecules are recaptured should be chosen such that the phase space volume of the molecules after expansion is matched to the acceptance of the trap~\cite{vandeMeerakker:ChemRev2012}. This implies that the optimal voltage used for recapturing the molecules actually depends on the expansion time.   

The ability to observe molecules after switching off the trap voltages for times $>$10\,ms is a promising prospect for future high-resolution experiments on molecules. The statistical uncertainty achieved in a spectroscopic experiment after a certain measurement time, $\tau$, is given by the Allan deviation

\begin{equation}
\sigma_y(\tau)=\frac{1}{Q}\sqrt{\frac{\tau_{c}}{\tau N_{c}}},
\end{equation}
where $Q=f/\Delta f$ is the quality factor of the resonance, $\tau_{c}$ is the duration of one cycle, i.e., the inverse of the repetition frequency and $N_{c}$ is the number of molecules that are detected per cycle. If the width of the transition is limited by the finite interaction time, $\Delta t$ (i.e., $\Delta f \propto 1/\Delta t$), then the Allan deviation is proportional to $(\Delta t \sqrt{N_{c}})^{-1}$. If we performed a coherent measurement during the time that the molecules are freely expanding, we would expect from the data presented in Fig.~\ref{fig4-recapturing} that the Allan deviation would decrease as $1/\sqrt{\Delta t_\mathrm{exp}}$. We can check this conclusion by explicitly computing some numbers; after an expansion time of 0.1,~1,~10 and 30\,ms, we detected (at the optimal voltage settings and corrected for background signal) 117,~18,~1, and 0.2 ions per shot, respectively. This implies that the Allan deviation at 1,~10 and 30\,ms is decreased by a factor of 4,~9 and 12, respectively, compared to the value at 0.1\,ms. Thus, the increased interaction time indeed compensates the loss in signal.
   
\section{Conclusion}

As shown in previous publications~\cite{Quintero-Perez2013, Jansen2013}, traveling-wave decelerators are more efficient at low velocity and obviate the problem of loading molecules into a trap, that has led to large losses in earlier studies with Stark decelerators~\cite{Bethlem:PRA2002,Gilijamse:EPJD2010}. Compared to the setup used in our previous experiments, we have made a number of improvements that have further reduced losses in the deceleration process. In the current experiments, the signal of molecules that are trapped for 50\,ms is only 20\% smaller than that of molecules that are guided by the traveling-wave decelerator. Furthermore the guided signal was 12 times larger than that in our previous experiments. These improvements have allowed us to lower the voltages of the trap by a factor of 50, which corresponds to lowering the trap depth from 160\,mK to 0.2\,mK, while still having 5\% of our original signal left. We have also performed experiments in which we switch off the trap completely and recapture the molecules more than 10\,ms later -- much longer than possible in a molecular beam machine of realistic size. We show that by performing adiabatic cooling before the molecules are released and adiabatic re-compression after the molecules are recaptured, the signal drops only as $1/\Delta t_\mathrm{exp}$. This implies that the statistical error of a spectroscopic experiment performed during that time will decrease as $1/\sqrt{\Delta t_\mathrm{exp}}$. Furthermore, the systematic errors are easier to control by the localised and well-controlled sample of cold molecules. The signal can be further increased by using larger rings, which will also improve the optical access. We are planning to perform precision measurements on the fundamental vibrations in ammonia using infrared lasers. As a first step, we plan to measure transitions in the $\nu_{1} + \nu_{3}$ band of NH$_{3}$ near 1.5\,$\mu$m \cite{Berden:CPL1999}.   

\section{Acknowledgements}
This research has been supported by NWO via a VIDI-Grant and by the FOM-program ``Broken Mirrors \& Drifting Constants''. We acknowledge the expert technical assistance of Leo Huisman, Imko Smid and Rob Kortekaas. We thank Paul Jansen and Wim Ubachs for helpful discussions.
 

\end{document}